\documentclass{PoS}
\usepackage{amsmath, amssymb}
\usepackage{axodraw}
\usepackage{psfrag}
\usepackage{graphicx}
\usepackage{epsfig}

\title{Kaons on the lattice}

\ShortTitle{Kaons on the lattice}

\author{\speaker{Christopher Sachrajda}%
         \\
        University of Southampton, School of Physics and Astronomy, Highfield, Southampton,\\ SO17 1BJ, United Kingdom\\
        E-mail: \email{cts@phys.soton.ac.uk}}

\abstract{I review recent lattice results in kaon physics, particularly in the determination of $V_{us}$ and the $B_K$ parameter of $K^0$-$\bar{K}^0$ mixing. I use lattice data to argue for the need of developing SU(2)$_L\times$SU(2)$_R$ chiral perturbation theory for kaon physics and discuss some recent progress in achieving this. In particular it is shown that for $K_{\ell 3}$ decays at $q^2=0$ (where $q$ is the momentum transfer between the kaon and the pion), the chiral logarithms can be calculated in spite of the fact that the external pion carries half the energy of the kaon (in the kaon's rest frame), because these logarithms arise from soft internal loops. Future prospects, including applications to $K\to\pi\pi$ decays are discussed. The need to define and exploit renormalization schemes which can simultaneously be implemented numerically in lattice simulations and used in higher-order perturbative calculations is explained.}

\FullConference{6th International Workshop on Chiral Dynamics, CD09\\
		July 6-10, 2009\\
		Bern, Switzerland}

\begin{document}
\section{Introduction}
It is a pleasure to have been invited to present lattice results on kaon physics at a conference on \textit{chiral dynamics}. In recent years the interactions between the Lattice QCD and Chiral Perturbation Theory (ChPT) communities have grown very significantly, and this collaboration is very important for the development of our understanding of non-perturbative QCD effects in flavour physics and hadronic structure. On the one hand, lattice simulations are performed at values of the up and down quark masses which are larger than the physical ones and ChPT is used to guide the extrapolation of the results to the physical point. On the other hand the ability to vary the masses of the quarks in lattice simulations allows us to compute the low-energy constants of ChPT with unprecedented precision. It is to be expected that this symbiotic relation will strengthen as the precision of both the lattice and ChPT calculations improves further.

The aims of this talk are i) to review recent results for important physical quantities including $f_K/f_\pi$, the form factors of semileptonic $K\to\pi$ decays and the $B_K$ parameter of $K^0$-$\bar{K}^0$ mixing; ii) to present a discussion of some of the theoretical conclusions, including the need to develop SU(2) ChPT for kaon physics and iii) to discuss future prospects including applications to $K\to\pi\pi$ decays. I start however, with some brief comments about the chiral behaviour of $f_\pi$. Many of the ideas presented below have been developed together with my colleagues from the RBC-UKQCD collaboration, and where appropriate I will illustrate the discussion with our results.

\section{Comments on Chiral Behaviour}
The ability to vary the masses of the quarks in lattice simulations allows us to study the chiral behaviour of physical quantities in considerable detail. Different collaborations use different approaches to including the NNLO (two-loop) and higher corrections, either fully or partially, coupling this with discussions of the lattice artefacts and finite volume corrections. There are a number of talks from different collaborations at this conference discussing their procedures in some detail~\cite{talks} and illustrating the dynamic activity in this field.

\begin{figure}[t]
\begin{center}
\includegraphics[angle=-90,width=0.3\hsize]{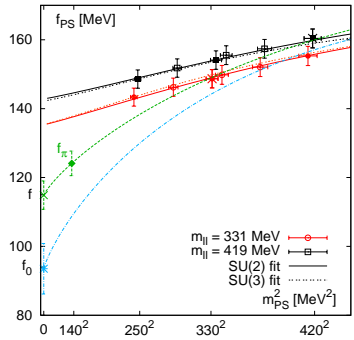}
\caption{Partially quenched study of the chiral behaviour of the pseudoscalar decay constant $f_{\textrm{PS}}$~\cite{Allton:2008pn}.\label{fig:fpichiral}}
\end{center}
\end{figure}

In this section I will focus on one important question, whether SU(3) and/or SU(2) ChPT adequately describe the lattice data. With my colleagues from the RBC-UKQCD collaboration, we argue that it is SU(2) ChPT which should be used, and I now summarise the argument. Fig.\,\ref{fig:fpichiral} shows the behaviour of pseudoscalar decay constant as a function of the mass of the meson. The right-hand upper (black) point marked with a cross corresponds to a unitary meson with $m_{\textrm{PS}}\simeq 420\,$MeV. The remaining upper (black) points have the same sea-quark mass but a variety of valence-quark masses. Similarly in the lower (red) band the point marked with a cross is unitary with $m_{\textrm{PS}}\simeq 330\,$MeV and the remaining points correspond to the same sea-quark mass but with different valence quark masses, keeping the meson mass below 420\,MeV. The data are fit to NLO Partially Quenched ChPT (PQChPT) and the dashed (green) curve represents the resulting unitary SU(2) behaviour, which, as expected, passes through the two unitary points. The same procedure is repeated with SU(3) PQChPT and the dot-dash (blue) curve represents the corresponding unitary behaviour with $m_s=m_{ud}$. The important point is that the resulting value for $f_0$ (the decay constant in the SU(3) chiral limit) is far below the data points (60-70\%), so that the validity of the expansion is questionable. It is for this reason that we advocate the use of SU(2) ChPT to study the chiral behaviour of lattice results (a further example of such a feature is shown in fig.\ref{fig:kpiextrap} discussed in sec.\ref{sec:kpipi}). Perhaps extending the SU(3) analysis beyond NLO may lead to an apparently more convergent series, and such studies are being performed but require more data at light masses and reliable techniques to separate chiral behaviour from small lattice subtleties. We therefore strongly prefer to present our results based on SU(2) ChPT and this is an area of active debate with the community.

\subsection{Kaon Chiral Perturbation Theory}\label{subsec:kaonchpt}
In order to apply SU(2) ChPT to kaons, the formalism must be extended. Roessl has introduced the corresponding Lagrangian for the interactions of kaons and pions so that he could study $K\pi$ scattering near threshold~\cite{Roessl:1999iu}. There are many similarities with \textit{heavy meson ChPT}~\cite{Wise:1992hn,Burdman:1992gh}, but an important difference is that in the heavy quark limit $m_{B^\ast}=m_B$, so that a $B$-meson can emit a soft pion (with coupling $g_{BB^\ast\pi}$) and turn into the vector meson $B^\ast$. The propagation of $B^\ast$ mesons must therefore be included in diagrams. This is not the case for kaons.

The NLO expressions for $m_K^2$, $f_K$ and $B_K$ in SU(2) ChPT (and partially quenched SU(2) ChPT) can be found in ref.\cite{Allton:2008pn} and in section \ref{subsubsec:su2chptkl3} I review the applications to $K_{\ell 3}$ decays. The applications of SU(2) ChPT to $K\to\pi\pi$ decays is discussed in the talk by Hans Bijnens based on ref.\,\cite{Bijnens:2009yr}.

\section{The Determination of $V_{us}$}

I start with a brief analysis within the standard model which I learned from my colleagues in the Flavianet Lattice Averaging Group (FLAG)~\cite{flag}. This provides a very useful benchmark for the lattice results presented below. Consider the two very precise experimental results~\cite{Antonelli:2008jg}:
\begin{enumerate}
\item From the ratio of the measured rates for leptonic decays of kaons and pions we have:
\begin{equation}\label{eq:lept}
\left|\frac{V_{us}\,f_K}{V_{ud}\,f_\pi}\right|=0.27599(59)\,,\end{equation}
where $f_K$ ($f_\pi$) is the decay constant of the $K$ ($\pi$) and contains the QCD effects in the decay. Thus a precise computation of $f_K/f_\pi$ will give an accurate result for $|V_{us}/V_{ud}|$.
\item From the measurement of the differential decay rate for semileptonic $K\to\pi$ decays ($K_{\ell 3}$ decays) we obtain
\begin{equation}\label{eq:semilept}
|V_{us}\,f_+(0)|=0.21661(47)\,,
\end{equation}
where $f_+(0)$ is one of the form factors for $K^-\to\pi^-\ell\nu$ decays at zero momentum transfer. Thus a precise computation of $f_+(0)$ will yield the value of $|V_{us}|$.
\end{enumerate}
In addition, within the standard model we have the unitarity relation
\begin{equation}\label{eq:unitarity}
|V_{ud}|^2+|V_{us}|^2=1\,,
\end{equation}
where $|V_{ub}|^2$ has been omitted since it is much smaller than the uncertainties on the left-hand side.

Eqs. (\ref{eq:lept})\,--\,(\ref{eq:unitarity}) can be viewed as 3 equations for the 4 unknowns $f_K/f_\pi,\,f_+(0),\,V_{ud}$ and $V_{us}$. We therefore only require one additional piece of information to determine these four quantities. One might take, for example, the recent determination of $V_{ud}$ based on 20 different
superallowed nuclear $\beta$-decays~\cite{Hardy:2008gy}
\begin{equation}\label{eq:vudhardy}
|V_{ud}|=0.97425(22)\,.\end{equation}
If we accept this value of $V_{ud}$, then we are able to determine the remaining 3 unknowns:
\begin{equation}\label{eq:threevars}
|V_{us}|=0.22544(95),\quad f_+(0)=0.9608(46),\quad \frac{f_K}{f_\pi}=1.1927(59)\,.\end{equation}

Of course, when exploring the limits of validity of the standard model we should not assume the unitarity relation (\ref{eq:unitarity}). I now describe lattice calculations of $f_K/f_\pi$ and $f_+(0)$.

\subsection{$V_{us}$ from $K_{\ell 2}$ Decays}\label{subsec:kl2}

\begin{figure}[t]
\begin{center}
\includegraphics[angle=-90,width=0.5\hsize]{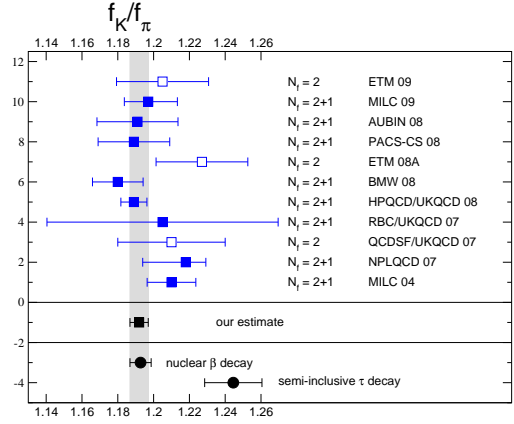}
\end{center}
\caption{A (preliminary) compendium of recent lattice results for $f_K/f_\pi$ from FLAG~\cite{flag}. See \cite{flag} for publication details of the individual contributions.\label{fig:fkfpi}}
\end{figure}

The difference of $f_K/f_\pi$ from 1 is an SU(3) breaking effects, and it is important to note that we calculate this difference rather than the ratio itself; in the SU(3) limit we would find precisely 1. The calculation relies on a good control of the chiral extrapolation. Recent results with two or three flavours of sea quarks are collected in fig.\ref{fig:fkfpi}. The results with small quoted errors include
\begin{equation}\label{eq:fkfpiresults}
\frac{f_K}{f_\pi}=1.197(3)\left(^{+6}_{-13}\right)\,\textrm{MILC}(09)\,\cite{Bazavov:2009bb}
\quad\textrm{and}\quad\frac{f_K}{f_\pi}=1.189(2)(7)\,\textrm{HPQCD/UKQCD}(08)\,
\cite{Follana:2007uv}\,.
\end{equation}
The preliminary FLAG summary value based on these results with $N_f=2+1$ sea-quark flavours is $f_K/f_\pi=1.190(2)(10)$\,\cite{flag}. This result is in remarkable agreement with that in eq.(\ref{eq:threevars}) which was obtained assuming the standard model and the value of $V_{ud}$ in (\ref{eq:vudhardy}).

\subsection{$V_{us}$ from $K_{\ell 3}$ Decays}\label{subsec:kl3}

In order to determine $V_{us}$ from $K_{\ell 3}$ decays we need a precise determination of the form factor $f_0(0)=f_+(0)$ (see eq.(\ref{eq:semilept})). The SU(3) chiral expansion for this form factor takes the form $f_+(0)=1+f_2+f_4+\cdots$, where $f_n=O(m_{\pi,K,\eta}^{2n})$. $f_2=$\,--\,$0.023$ contains no low energy constants and is well determined. $f_4$ and higher order coefficients require additional theoretical or model input and a benchmark value is that obtained by Leutwyler and Roos in 1984, $f_+(0)=0.961\pm0.008$~\cite{Leutwyler:1984je}. We see therefore that in order to be useful in extracting $V_{us}$, we need to be able to compute $f_+(0)=f_0(0)$ to better than about 1\% precision. The apparently challenging target is possible because again it is the SU(3) breaking effects which we actually compute; the form factor is precisely 1 in the SU(3) limit and it is the difference from 1 which we calculate.

\begin{table}
\begin{center}
\begin{tabular}{|cccc|}\hline
\rule[-0.2cm]{0cm}{0.55cm}$am_{ud}$&$m_\pi$& $q^2_{\textrm{max}}$
(GeV$^2$)&$f_0(q^2_{\textrm{max}})$\\ \hline
\rule[-0.2cm]{0cm}{0.55cm}0.03&670\,MeV&
0.00235(4)&1.00029(6)\\
\rule[-0.2cm]{0cm}{0.55cm}0.02&555\,MeV&0.01152(20)&1.00192(34)\\
\rule[-0.2cm]{0cm}{0.55cm}0.01&415\,MeV&0.03524(62)&1.00887(89)\\
\rule[-0.2cm]{0cm}{0.55cm}0.005&330\,MeV&
0.06070(107)&1.02143(132)\\
\hline
\end{tabular}\end{center}
\caption{The values of the form factor $f_0(q^2_{\textrm{max}})$ at the four quark masses corresponding to the four pion masses given in the second column~\cite{Boyle:2007qe}.\label{tab:f0qsqmax}}
\end{table}

We start with the calculation of the form-factor $f_0(q^2_{\textrm{max}})$, where $q^2_{\textrm{max}}=(m_K-m_\pi)^2$ and corresponds to the kaon and pion at rest in the same frame~\cite{Becirevic:2004ya}. By calculating the ratio
\begin{equation}
\frac{\langle\pi|\bar{s}\gamma_4 u|K\rangle
\langle K|\bar{u}\gamma_4 s|\pi\rangle}{\langle\pi|\bar{u}\gamma_4
u|\pi\rangle \langle K|\bar{s}\gamma_4
s|K\rangle}=\left[f_0(q^2_{\textrm{max}})\right]^2\,\frac{(m_K+m_\pi)^2}{4m_Km_\pi}\,,
\end{equation}
we can obtain $f_0(q^2_{\textrm{max}})$ with
excellent precision. This is illustrated by the RBC-UKQCD results~\cite{Boyle:2007qe} for $f_0(q^2_{\textrm{max}})$ in table\,\ref{tab:f0qsqmax}.

Having obtained $f_0(q^2_{\textrm{max}})$ at the quark masses used in the simulation, we need to interpolate to $q^2=0$ and extrapolate to the physical masses. Conventionally the $q^2$ interpolation is done by calculating the form factors with one of the mesons at non-zero momenta and the results presented below are obtained in this way. It is also possible to evaluate the form factors directly at $q^2=0$ by using partially twisted boundary conditions~\cite{Boyle:2007wg,Flynn:2008hd}.

The chiral extrapolation is more problematical and we would welcome all the guidance which the chiral dynamics community is able to provide. The SU(3) ChPT prediction at NLO is $f_0(0)=1+f_2$ which both the RBC-UKQCD\,\cite{Boyle:2007qe} and ETMC\,\cite{Lubicz:2009ht} collaborations find to lie considerably above their lattice results. It is not possible to exploit the two-loop results from ref.\,\cite{Bijnens:2003uy}, since the results are expressed as a series with the physical pion decay constant in the expansion parameter and coefficients given in terms of integrals to be evaluated numerically. The NLO SU(2) ChPT result for $f_0(0)$ was derived in\,\cite{Flynn:2008tg} and is reproduced in eq.(\ref{eq:f0su2qsq0}) below, but in order to use it we would like  results at more values of the light-quark mass to confirm that the chiral regime is indeed reached. The final results from the two collaborations are
\begin{equation}\label{eq:kl3results}
f_0(0)=0.964(5)\quad\textrm{RBC-UKQCD \cite{Boyle:2007qe};}\qquad f_0(0)=0.9560(57)(62)
\quad\textrm{ETMC \cite{Lubicz:2009ht}}\,.
\end{equation}
These values are lower than those obtained using analytical methods based on the two-loop ChPT formulae of\,\cite{Bijnens:2003uy}.

\begin{figure}
\begin{center}
\includegraphics[width=0.3\hsize,angle=270]{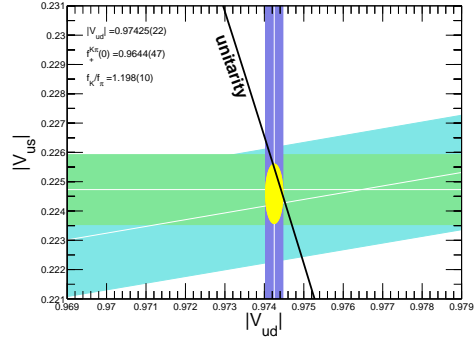}%
\caption{Determination of $V_{us}$ and $V_{ud}$ based on the lattice results for $V_{us}/V_{ud}$
and the RBC-UKQCD result for $V_{us}$.
The vertical blue band corresponds to $V_{ud}$ in~\cite{Hardy:2008gy}.
The black curve is the unitarity condition $|V_{ud}|^2+|V_{us}|^2=1$\,.\label{fig:vusvud4}}
\end{center}
\end{figure}

Comparing the lattice results in eqs.(\ref{eq:fkfpiresults}) and (\ref{eq:kl3results}) with the expectations from the standard model~(\ref{eq:threevars}) (which, it should be noted, was based on the value of $V_{ud}$ in\,(\ref{eq:vudhardy})) we see that there is very little room for new physics contributions to the violation of the unitarity of the first row of the CKM matrix. This is illustrated in fig.\,\ref{fig:vusvud4}~\footnote{I am grateful to Andreas J\"uttner for providing this figure and to the Flavianet Kaon Working Group for developing this way of presenting the results.} and will be quantified in\,\cite{flag}.

\subsubsection{SU(2) Chiral Perturbation Theory for $K_{\ell 3}$ Decays}
\label{subsubsec:su2chptkl3}
In this section I review the applications of SU(2) chiral perturbation theory to $K_{\ell 3}$ decays at $q^2=q^2_{\textrm{max}}$ where it is natural and at $q^2=0$ where, because the pion is hard, it is not~\cite{Flynn:2008tg}.

\paragraph{$K_{\ell 3}$ decays at $q^2_{\textrm{max}}$.}
SU(2) chiral perturbation theory can naturally be applied to $K_{\ell 3}$ decays near the end point of phase space where the momentum of the pion is small (in the rest frame of the kaon). Although we cannot perform simulations at the unphysical Callan-Treiman point where $q^2=m_K^2-m_\pi^2$, as $m_u=m_d\to 0$ the value of $q^2$ at the edge of the physical phase space, $q^2_{\textrm{max}}=(m_K-m_\pi)^2$ approaches the Callan-Treiman point and we have the relation
\begin{equation}\label{eq:ct}
f_0(q^2_{\textrm{max}})\ \underset{m_{\pi}\to 0}\longrightarrow\ \frac{f^{(K)}}{f}\simeq 1.26\,,
\end{equation}
where $f^{(K)}$ and $f$ are the kaon and pion decays constants in the SU(2) chiral limit, and the value 1.26 is estimated on the basis of the chiral behaviour of the ratio of decay constants in the RBC-UKQCD simulation~\cite{Allton:2008pn}~\footnote{Thanks to Enno Scholz for providing this number.}. On the other hand when we look in table\,\ref{tab:f0qsqmax} at the RBC-UKQCD values of $f_0(q^2_{\textrm{max}})$ obtained with pion masses ranging from 670\,MeV to 330\,MeV~\cite{Boyle:2007qe} there is little indication of convergence towards 1.26 in the chiral limit. We investigated this using $SU(2)$ ChPT and found that although the chiral logarithms are of approximately the correct size to account for the difference between the results in table\,\ref{tab:f0qsqmax} and 1.26 they have the wrong sign, tending to make the form factor decrease towards the chiral limit. The result is
\begin{equation}\label{eq:f0qsqmax}
f_0(q^2_{\textrm{max}})=\frac{f^{(K)}}{f}\,\left[1-\frac{11}{4}\,\frac{m_\pi^2}{(4\pi f)^2}\log\frac{m_\pi^2}{\Lambda_\chi^2}+\frac{\lambda_1}{4\pi f}\,m_\pi+
\frac{\lambda_2}{(4\pi f)^2}\,m_\pi^2+\cdots\right]\,.
\end{equation}
The coefficients $\lambda_{1,2}$ cannot be evaluated within $SU(2)$ ChPT. They can be estimated however, by converting the results from $SU(3)$ chiral perturbation theory, where there are no unknown constants at one-loop level except for $f^{(K)}/f$~\cite{Gasser:1984ux,Gasser:1984gg}. This conversion suggests that they have the correct sign and magnitude to account for the difference of the observed results from 1.26.

It should be noted that this rapid chiral behaviour at the edge of phase space is not limited to the form factor of the kaon, but also applies to the heavy mesons where
\[
f_0^{D\to\pi}(q^2_{\textrm{max}})\underset{m_\pi\to 0}{\longrightarrow}\frac{f^{(D)}}{f}\qquad\textrm{and}\qquad
f_0^{B\to\pi}(q^2_{\textrm{max}})\underset{m_\pi\to 0}{\longrightarrow}\frac{f^{(B)}}{f}\,,
\]
and yet in the region of quark masses where we have data, the $f_0(q^2_{\textrm{max}})$ are below the expected ratios of decay constants. As an example consider fig.14 in ref.\cite{Dalgic:2006dt} where the form factor $f_0$ for $B\to\pi$ semileptonic decays does not appear to approach the expected value of $f^{(B)}/f$ as $q^2\to q^2_{\textrm{max}}$.

\paragraph{$K_{\ell 3}$ decays at $q^2=0$.} At first sight it appears that SU(2) ChPT should not be applied at $q^2=0$, since in this case the final-state pion is hard; in the rest frame of the kaon its energy is $m_K/2$. However, we argue in \cite{Flynn:2008tg} that the chiral logarithms come from soft internal loops and that it is possible to evaluate them:
\begin{eqnarray}\label{eq:f0su2qsq0}
f_0(0)=f_+(0)&=&F_+\left(1-\frac34 \frac{m_\pi^2}{16\pi^2f^2}\log\left(\frac{m_\pi^2}{\mu^2}\right)+c_+m_\pi^2\right)\\
f_-(0)&=&F_-\left(1-\frac34 \frac{m_\pi^2}{16\pi^2f^2}\log\left(\frac{m_\pi^2}{\mu^2}\right)+c_-m_\pi^2\right)\,,
\label{eq:f-su2qsq0}\end{eqnarray}
where $F_{+,-}$ and $c_{+,-}$ are unknown constants which depend on $m_s$ but not on $m_{ud}$. In arriving at (\ref{eq:f0su2qsq0}) and (\ref{eq:f-su2qsq0}) we use integration by parts and equations of motion to reduce the infinite chain of \textit{higher order} operators which contain additional derivatives on the hard external pion to the leading operator. It may be possible to formulate this approach to ChPT with hard external pions in terms of an effective theory in which hard and soft pions are separated, but this has not yet been achieved.

The approach described here can also be applied to the semileptonic form factors at other values of $q^2$ and to other processes. It has been applied to nonleptonic kaon decays~\cite{Bijnens:2009yr} and in the preceding talk Bijnens has referred to the approach as \textit{Heavy Pion Chiral Perturbation Theory}.

Since the chiral extrapolation is the major source of systematic uncertainty for the lattice determination of $V_{us}$ from $K_{\ell 3}$ decays, it is important to have all the possible theoretical information to guide us, and in particular it would be very helpful to extend the result in (\ref{eq:f0su2qsq0}) to NNLO.

\section{$B_K$}\label{sec:BK}
\begin{figure}
\begin{center}
\includegraphics[width=0.5\hsize]{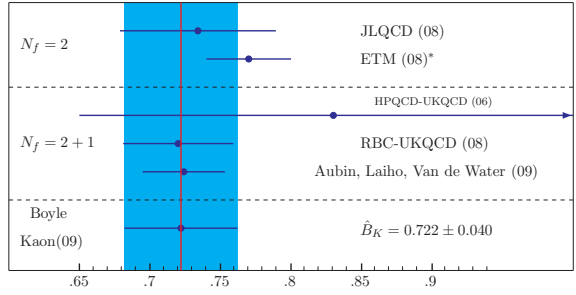}
\end{center}
\caption{Recent unquenched results for $\hat{B}_K$ together with the average proposed by P.Boyle at the 2009 Kaon International Conference.\label{fig:bkrecent}}
\end{figure}

Calculations of the parameter $B_K$ which contains the non-perturbative QCD effects in $K^0$\,--\,$\bar{K}^0$ mixing have been performed for about 20 years. $B_K$ is defined in terms of the matrix element of the single $\Delta S=2$ operator which contributes in the Standard Model:
\begin{equation}\label{eq:bkdef}
\langle\bar{K}^0\,|\,(\bar{s}\gamma_\mu(1-\gamma_5)d)\,
(\bar{s}\gamma^\mu(1-\gamma_5)d)\,|K^0\rangle
=\frac83\,f_K^2 m_K^2 B_K(\mu)\,.\end{equation}
If the lattice formulation has good chiral and flavour symmetries then the $\Delta S$=2 operator in (\ref{eq:bkdef}) renormalizes multiplicatively. The renormalization can be performed nonperturbatively. Recent unquenched results for
the scale invariant parameter $\hat{B}_K$ are shown in fig.\,\ref{fig:bkrecent}.
I highlight two calculations with results which are in remarkable agreement. The first is the RBC-UKQCD result presented in~\cite{Antonio:2007pb,Allton:2008pn}, $\hat{B}_K=0.720(13)(37)$ which was obtained from its domain wall fermion (DWF) datasets at a single lattice spacing but with an estimate of 4\% for the discretisation uncertainties; this is the largest component of the quoted error. ($\hat{B}_K=0.720(13)(37)$ corresponds to $B^{\overline{\textrm{MS}}}_K(2\,\textrm{GeV})=0.524(10)(28)$.) The second calculation by Aubin, Laiho and Van de Water obtained $\hat{B}_K=0.724(8)(28)$~\cite{Aubin:2009jh} and was obtained using domain wall valence quarks and staggered sea quarks at two lattice spacings. The \textit{taste} unitarity violations were removed using SU(3) CHPT at NLO (including some of the NNLO terms). In both cases the $\Delta S=2$ operator was renormalized non-perturbatively using the same techniques.

Given these results I am happy to take
\begin{equation}\label{eq:bkbest}
\hat{B}_K=0.722(40)
\end{equation}
as the current best value as was proposed by P.Boyle at the 2009 Kaon International Conference.

\section{$K\to\pi\pi$ Decays}\label{sec:kpipi}

An understanding of the longstanding puzzle of the $\Delta I=1/2$ rule and the theoretical determination of $\varepsilon^\prime/\varepsilon$, whose experimental measurement confirmed the existence of direct CP-violation in kaon decays, are major long-term goals for the lattice community. In 2001, two collaborations published some very interesting quenched results on non-leptonic kaon decays, in
particular,
\begin{center}
\begin{tabular}{|c|c|c|}\hline Collaboration(s)&{\tt Re}
$A_0$/{\tt Re} $A_2$ & $\varepsilon^\prime/\varepsilon$\\ \hline
RBC& $25.3\pm 1.8$ &$-(4.0\pm 2.3)\times 10^{-4}$\\
CP-PACS
& 9$\div$12&(-7\,$\div$\,-2)$\times 10^{-4}$\\
Experiments & 22.2 & $(17.2\pm 1.8)\times 10^{-4}$\\
\hline
\end{tabular}\end{center}
where $A_0$ and $A_2$ are the decays into isospin $I=0$ and $I=2$ two-pion final states. These results were obtained not only in the quenched approximation, but also at lowest order in the $SU(3)$ chiral expansion and at relatively heavy masses ($m_\pi> 400-500$\,MeV). At Lowest Order in SU(3) ChPT the $K\to\pi\pi$ decay amplitude is given in terms of $K\to\pi$ and $K\to$\,vacuum matrix elements, i.e. with at most one hadron in the initial and final state. In spite of the limitations of these calculations, the authors did achieve the control of the \textit{ultraviolet} problem, i.e. the numerical subtraction of power divergences and the renormalization of the weak operators. This is highly non-trivial.

\begin{figure}
\begin{center}
\includegraphics[width=0.5\hsize]{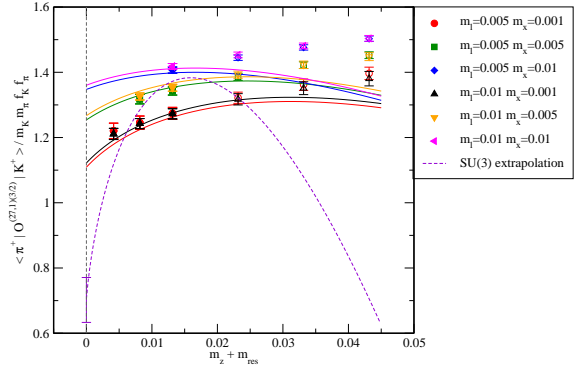}%
\end{center}
\caption{The mass dependence of the $K\to\pi$ matrix element. $m_l$, $m_x$ and $m_z$ are the masses of the sea and valence light quark and the valence "heavy" (strange) quarks respectively.\label{fig:kpiextrap}}
\end{figure}

The RBC/UKQCD collaboration have repeated the calculation with the $24^3$ DWF ensembles in the pion-mass range 240-420\,MeV~\cite{Li:2008kc,norman}. For illustration consider the determination of $\alpha_{27}$, the LO LEC for the $\Delta I=3/2$ (27,1) operator:
\begin{equation}
O^{3/2}_{(27,1)}=(\bar{s}d)_L\,\big\{(\bar{u}u)_L-(\bar{d}d)_L\big\}+(\bar{s}u)_L\,(\bar{u}d)_L\,,
\end{equation}
where the subscript $L$ stands for left. Satisfactory fits for the mass dependence were obtained using NLO SU(3) ChPT, but again the corrections were found to be very large, casting serious doubt on the approach. This is illustrated in fig.\ref{fig:kpiextrap}, the dashed curve represents the curve in the SU(3) limit ($m_u=m_d=m_s$) and the value of this curve in the chiral limit is much below the data points, demonstrating that the one-loop corrections are very large. Thus the use of soft-pion theorems is not sufficiently reliable and $K\to\pi\pi$ matrix elements have to be computed directly.

\begin{figure}
\begin{center}
\includegraphics[width=0.25\hsize]{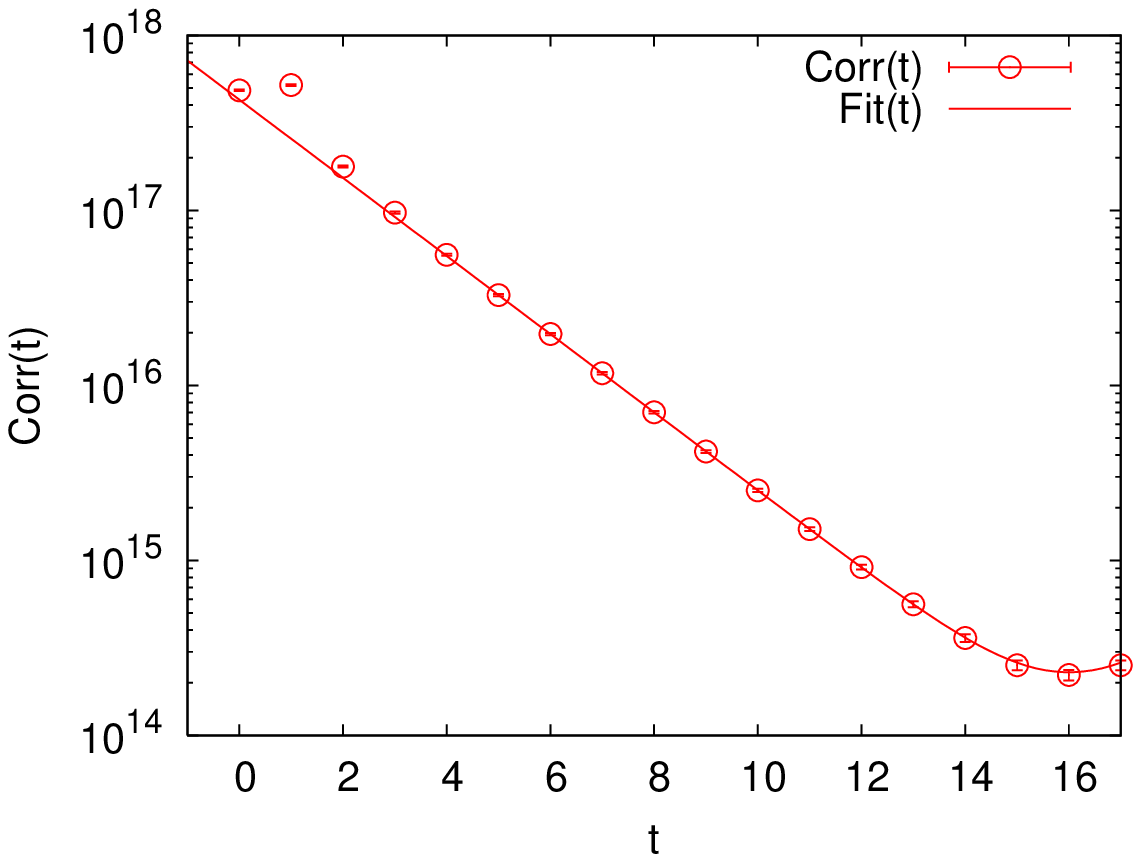}\hspace{0.3in}
\includegraphics[width=0.25\hsize]{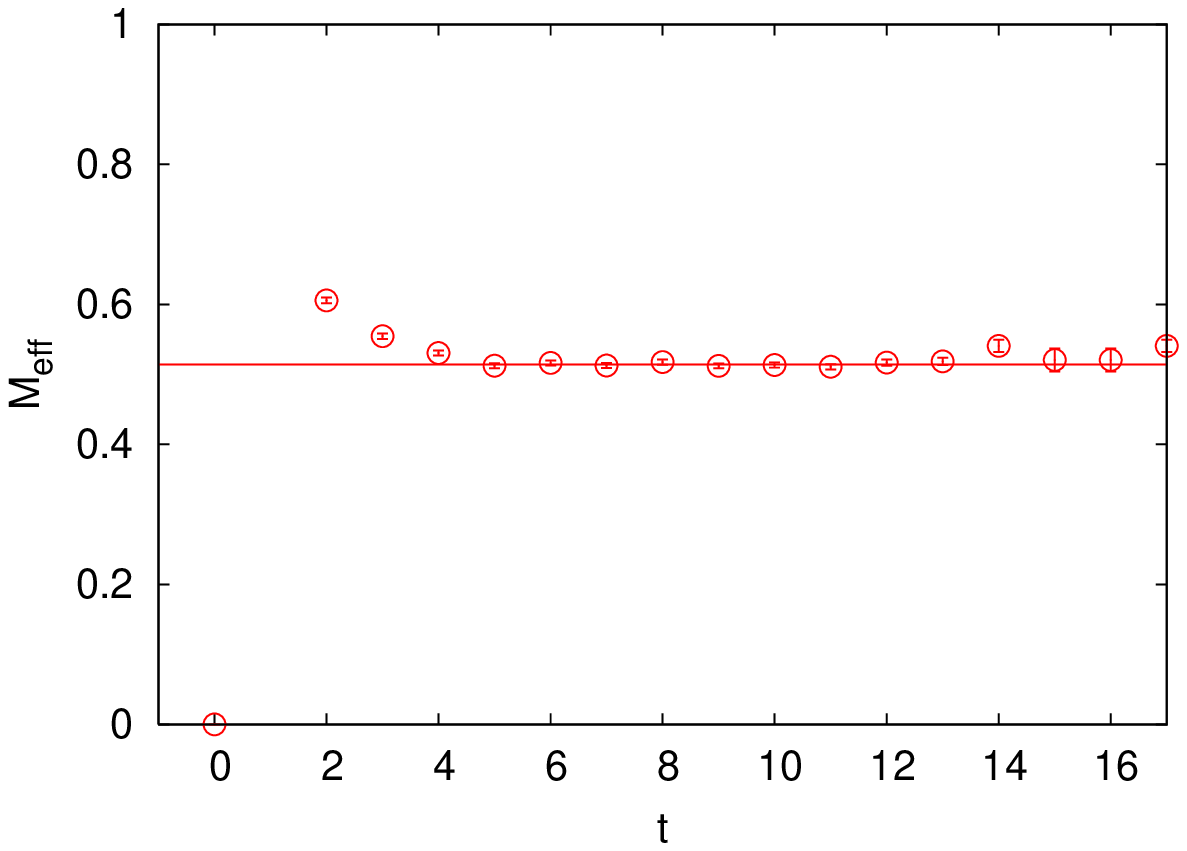}\\
\includegraphics[width=0.25\hsize]{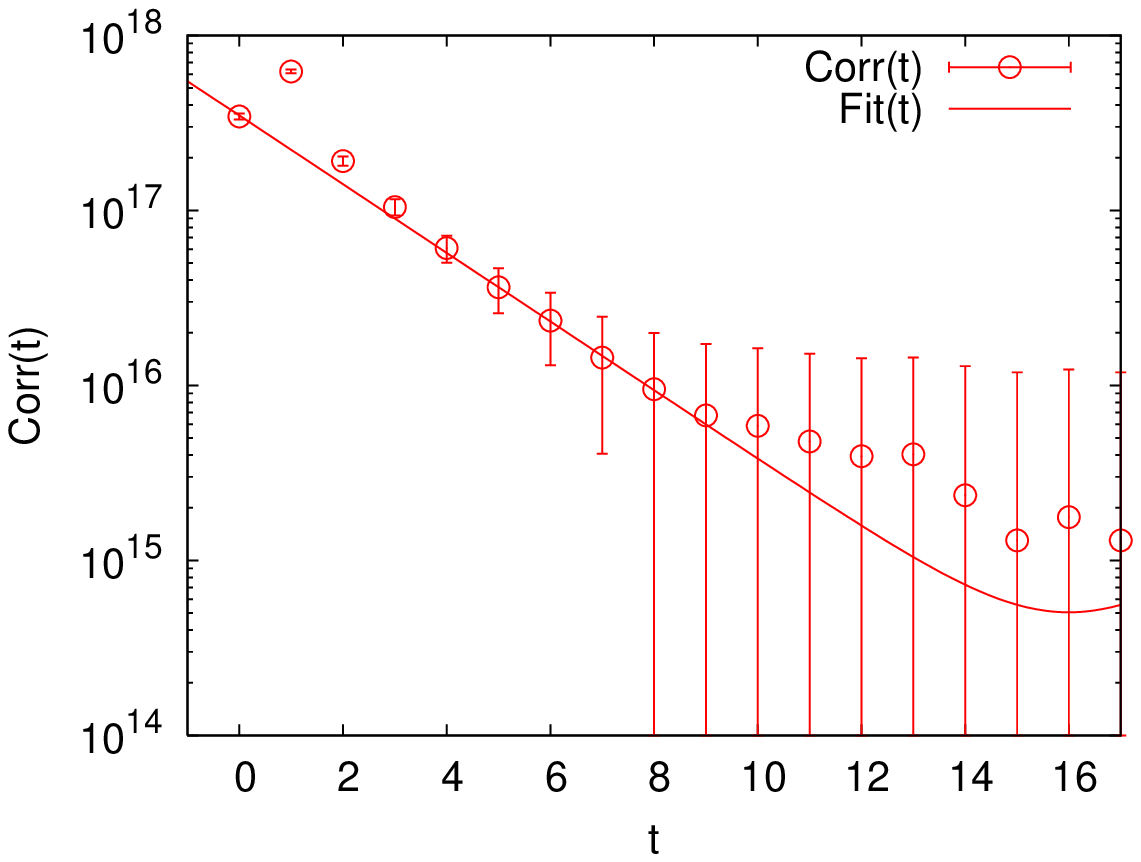}\hspace{0.3in}
\includegraphics[width=0.25\hsize]{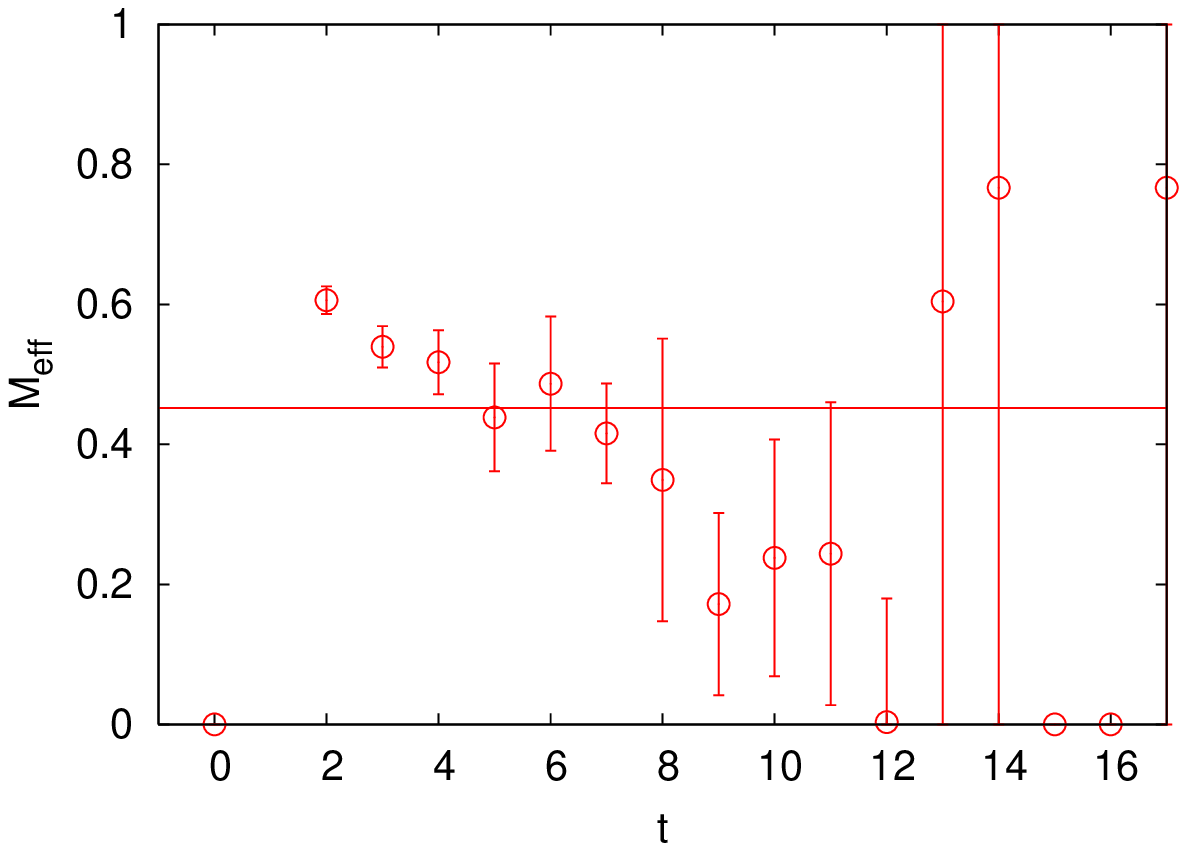}\\
\includegraphics[width=0.25\hsize]{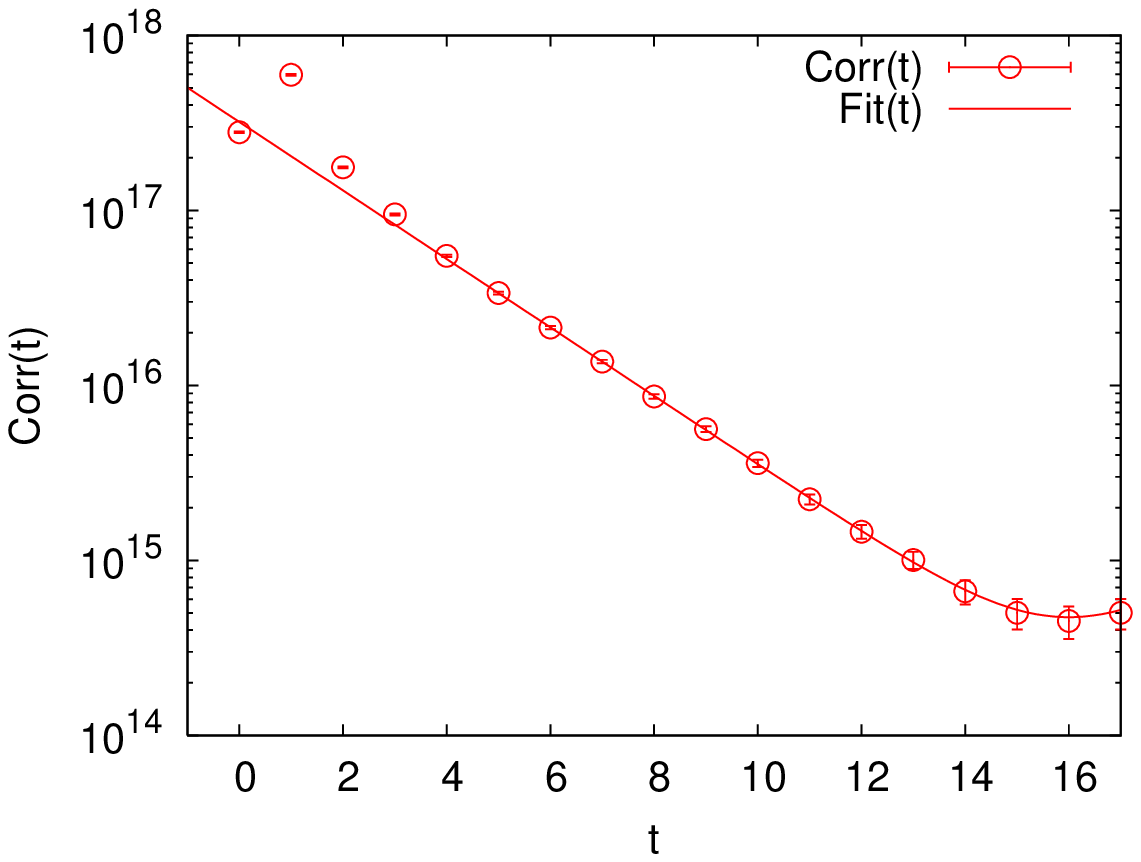}\hspace{0.3in}
\includegraphics[width=0.25\hsize]{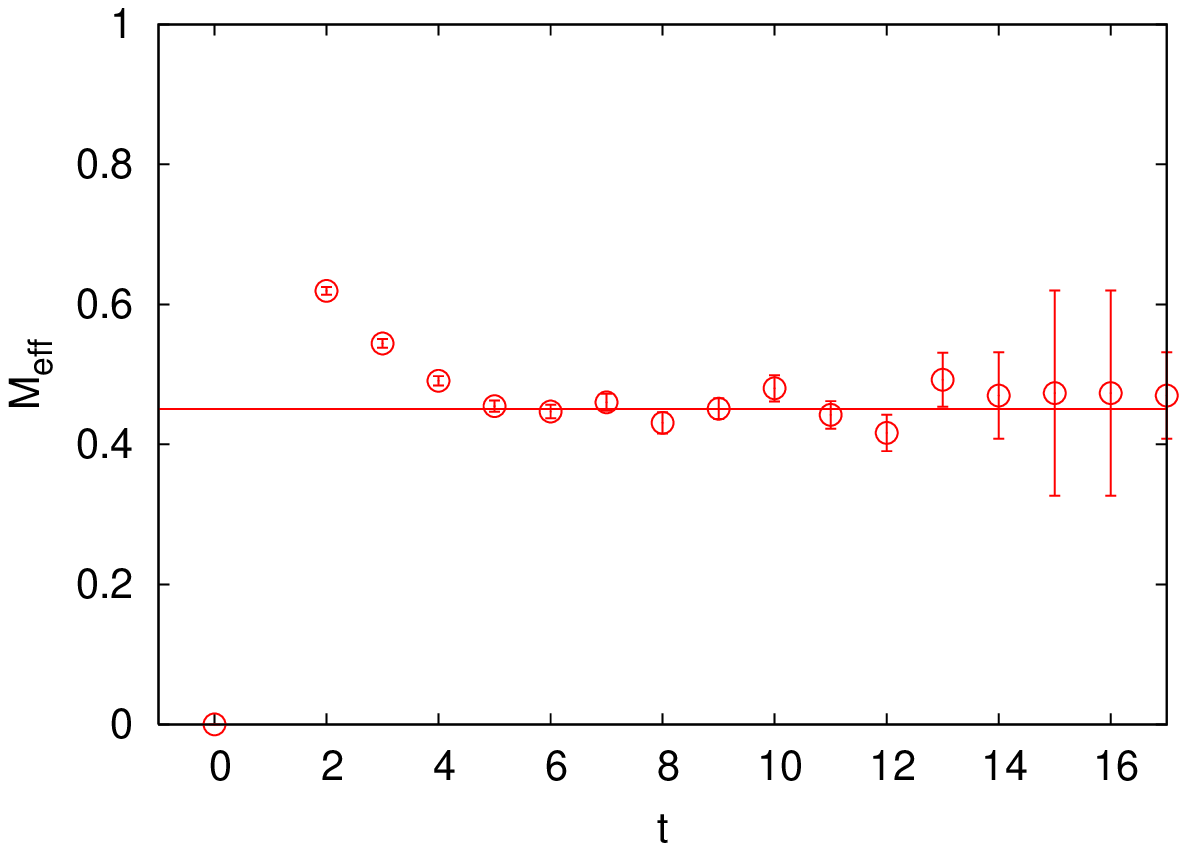}
\end{center}
\caption{The correlation function and effective mass for the $(\pi\pi)_{I=2}$ propagator (top line) and for the $(\pi\pi)_{I=0}$ propagator (middle line) as a function of time $t$. The bottom line shows the $(\pi\pi)_{I=0}$ propagator with the contribution from diagram $V$ removed. The data is from an exploratory calculation by RBC-UKQCD (courtesy of Qi Liu).\label{fig:pipi2}}
\end{figure}

In general the computation of matrix elements with multi-hadron states is not understood theoretically. For $K\to\pi\pi$ decays however, under the assumption that the s-wave phase-shift dominates the rescattering and that inelastic states can be neglected the theoretical framework is understood. Consider first the propagation of two pions, between time 0 and t. Four diagrams contribute to the two-pion propagator:
\begin{center}
\includegraphics[width=0.7\hsize]{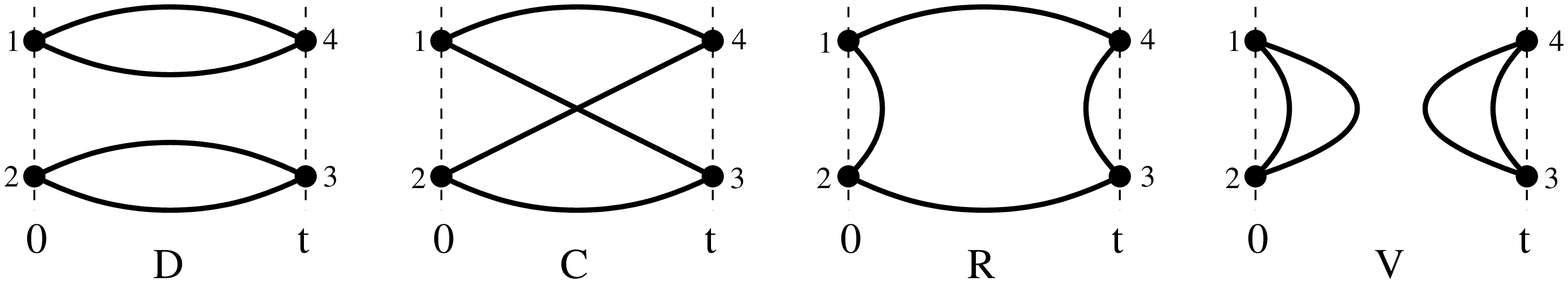}
\end{center}
If the two pions are in an I=2 state then only diagrams D and C contribute and the correlation function is proportional to D-C. For the propagation of the I=0 state all four diagrams contribute and the correlator is proportional to 2D+C-6R+3V. The principle difficulty in the evaluation of $K\to(\pi\pi)_{I=0}$ decay amplitudes is the evaluation of diagrams with a vacuum contribution such as diagram $V$ in the propagation of two pions. To illustrate this point consider the plots shown in fig.~\ref{fig:pipi2} which show the correlation functions for two pions (left-hand panels) and the corresponding effective masses (right-hand panels), both as functions of the time $t$. The data is from an exploratory study by the RBC-UKQCD collaboration and I am grateful to Qi Liu for providing them. The top row contains the results for the $I=2$ state and as expected these are very clean. The middle row on the other hand, shows the corresponding results for the $I=0$ state and we see that large errors start at an early time. The bottom row shows the $I=0$ correlation function but with the $V$ diagram removed, highlighting that the problem is largely due to the precision with which we can do the vacuum subtraction from the correlation function. In order to make further progress we need to try to develop techniques which would allow us to extend the range of $t$ where the errors are small and to use this range most effectively. Attempts to do this are under way.

\section{Conclusions and Prospects}
These are exciting times. Unquenched lattice simulations are presenting results with good precision with very light quarks and much effort is now being devoted to considering how lattice calculations will best help to clarify any new physics discovered at the LHC. The consistency of results with different actions is impressive and important and adds hugely to our confidence. The chiral regime is being mapped out for the spectrum and decay constants with some simulations bring performed with light quarks close to the physical ones. The more expensive simulations with good chiral and flavour properties, such as the domain wall fermions which were used by RBC-UKQCD in obtaining many of the results presented above, will be used to extend the range of physical quantities which can be studied in lattice kaon physics. I imagine that by the next Chiral Dynamics Workshop in 2012, the chiral behaviour of many fundamental quantities will be well understood and the low energy constants will be determined with excellent precision. For this to be achieved, there will have to be continued close collaboration between the ChPT and Lattice communities.

In this talk I have not had the opportunity to discuss how we perform the renormalization of bare lattice quantities or operators. I want to stress however, that the precision of lattice calculations has now reached the point where we need significant cooperation from the perturbative QCD community. Many (but not all) lattice calculations are implemented using \textit{non-perturbative renormalization}, in which the renormalization conditions are performed non-perturbatively and perturbation theory using the lattice action is completely avoided. For this to be possible the renormalization conditions have to be ones which can be simulated, such as ones based on the evaluation of Green functions at external momenta which serve as the renormalization scale~\cite{Martinelli:1994ty} or on the use of the Schr\"odinger functional~\cite{Jansen:1995ck}. Higher order perturbative calculations on the other hand tend to be performed using dimensional regularization which cannot be simulated. Thus, in order to combine lattice results for operator matrix elements with perturbatively computed Wilson coefficient functions, a matching between two continuum schemes must be performed in perturbation theory. The matching factor is frequently only available at one-loop order, leading to a significant error by today's standard (an important example is the evaluation of $m_s$ with domain wall fermions as discussed in\,\cite{Sturm:2009kb}, where the error in the determined mass in the $\overline{\textrm{MS}}$ scheme is largely due to this \textit{continuum} calculation.) Cooperation between the perturbative QCD and lattice communities to define renormalization schemes which can be implemented in lattice calculations and yet which are convenient for higher order perturbative calculations is now necessary.

I hope that in this talk I have demonstrated that a huge amount has already been achieved in lattice kaon physics, but that major challenges still lie ahead.

\paragraph{Acknowledgements} I warmly thank my colleagues from the RBC-UKQCD and FLAG Collaborations with whom I have developed much of my understanding of the material of the talk. I acknowledge partial support from STFC Grant ST/G000557/1 and EU contract MRTN-CT-2006-035482 (Flavianet).

\end{document}